***From Structure to Function in Open Ionic Channels***

May 24, 1999

Bob Eisenberg
Dept. of Molecular Biophysics and Physiology
Rush Medical Center
1750 West Harrison Street
Chicago Illinois 60612
USA

*Journal of Membrane Biology* **1999**, *171*, 1-24.

*Submitted to arXiv November 12, 2010 with text unchanged*

***Introduction.*** Life is diverse and complex in both structure and function. The variety of animals and structures within animals has been obvious at least since the time of Aristotle (Aristotle, 1961), and so has been the richness of what they can do. For more than two millennia scientists have followed Aristotle's path, trying to understand how structure produces function in biological systems, continually looking at smaller and smaller parts of the systems, trying to make 'the secrets of life' understandable, and controllable.

Progress along this path has been frustrating scientists for centuries. Every structure seems to be followed by still smaller structures, all important to natural function. But the end of the path can now be seen. Structures smaller than atoms are not directly involved in life's work, except in so far as electrons carry current and protons control the chemical properties of dissolved molecules. The role of macroscopic quantum coherence in the biological world has intrigued many (Loewenstein, 1999) but is not yet established. The smallest length scale directly relevant to life is that of molecules and their atoms. The magnificent tools of molecular biology make life's machines (proteins) and blueprints (nucleic acids) experimentally accessible. The machines and blueprints are on the molecular and atomic scale, and not on the length scale of the nucleus, the quark, or (fortunately) the electron.

The other part of the biologists' quest is to understand how these structures produce function. There, the goal is not in sight yet, although we will argue later in this paper that it may be fairly soon for one type of protein, ionic channels, that have simple structure when open, and use particularly simple physics, that of electrodiffusion.

Understanding the function of any biological systems means understanding how biological systems use physical laws to perform that function. When the biological systems consist of hierarchy upon hierarchy of structures, each itself of considerable complexity, the role of physical laws may be hard to recognize, at least in the form they are used by chemists and physicists. But open ionic channels have such simple structure, they involve so few hierarchical levels in their biological function, that we may be able to understand and solve them in the not too distant future. Fortunately, channels are of great biological importance, so despite their simplicity, they are worth studying.

Before we turn to channels explicitly, I will try the reader's patience (or ask the impatient reader to skip ahead to the section labeled ***Working Hypothesis***) with some more philosophical remarks about biological complexity, that are meant to show that not all biological systems use physical laws in the simple way they are used by open channels.

***Vitalism and Complexity***. Hierarchies can and do have qualitatively different properties from their components. The operation of an automobile engine cannot be understood just from the study of the burning of gasoline. The function of an integrated circuit or even transistor cannot be understood solely from the physics of conduction of current by quasi-particles. The nervous system cannot be understood from the physics of ionic conduction. In each case, knowledge of structure is needed as well as knowledge of underlying physics. The wiring diagram of the devices is as important as their physics.

The structure and underlying physics are not always enough to understand biological systems of complexity, because the complexity itself adds qualitatively new behaviors not evident in the underlying pieces of the system. While these new behaviors are certainly compatible with the underlying physical laws of the pieces, and in that sense implicit in them, they cannot be uniquely predicted from those underlying laws without a detailed understanding of the relevant hierarchy of structure. In many fewer words: a machine does much more than its parts do separately because its parts are designed to work together to perform a function.

When confronted with biological behaviors for which there is no technological precedent, like the speed with which the human visual system recognizes loved ones in a rain or snowstorm, it is sensible to seek explanations that are not well precedented in chemistry and physics, simply because chemical and physical systems have no such behaviors. It is sensible to seek explanations that arise from the hierarchy of structural complexity. In this quite limited sense, explanations are needed for biological systems that lie outside the laws of physics, as they are usually presented. The explanation must include both the physics and the structure, and, in a certain sense, the purpose of the structure, but it cannot consist only of the structure or only of the physics, at least in my view.

In this quite limited sense, then, vitalism is an appropriate part of biology. Physical laws undoubtedly govern the behavior of these complex systems, as well as governing the behavior of their elements, but, taken as a whole, biological systems, and organisms, often show behaviors that reflect the hierarchies of structures more than the properties of the elements of those structures. Those behaviors might be called 'vital', organic to the complexity of the structure, not obvious in the underlying physical laws.

Of course, this need to study complexity in its own right is not confined to biological systems. It is precisely what faces an engineer trying to understand a complex machine, if she has no clues about what it does or how it was built, other than those present in the machine itself. Thus, the word 'vitalism', which I used above is somewhat inappropriate, a piece of artistic license, that I hope may be granted me, with a smile on the reader's face.

***Vitalism and molecules***. When we confront the molecules of life, these semi-philosophical issues evaporate. The behaviors of individual molecules are much more closely related to physical laws than behaviors of complex biological systems or organisms. Every molecular biologist I have ever met agrees that vitalism is inappropriate in his or her science: explanations of the behavior of proteins and nucleic acids should be found in the laws of physics and chemistry, not in laws that describe complex biological systems.

When molecular biologists say they are not vitalists, more is being said than is often heard, even by the speakers themselves. The molecular biologists are in fact more or less disqualifying themselves as creators and even judges of the laws they will use to describe and analyze their molecules.

Just as the biologist is not responsible for the operating system of his computer, or the electronics of his oscilloscope or other instruments (with notable exceptions: Hodgkin, Huxley & Katz, 1952; Levis & Rae, 1992; Levis & Rae, 1995; Nonner, 1969; Sigworth, 1995; Valdiosera, Clausen & Eisenberg, 1974), so the biologist is not responsible for the physics and chemistry of the ionic solutions that his molecules live in. Molecular biologists must use physical laws as they are given us by our colleagues who study the physics and chemistry of ionic solutions. If those colleagues are successful, they will give us succinct

'laws' (often nowadays, computer programs) that summarize masses of experiments, covering the full range of conditions biologists need to describe.

These issues seem not very controversial, to me, when applied to proteins in general, and certainly to open ionic channels, but it is possible that a touch of vitalism is needed when describing gating properties of channels. It possible that the conformation changes that occur in channels involve a complexity in behavior not easily captured in ordinary physical language, even with the large number of states currently in fashion (more than 50). I believe that a proper understanding of the mechanical, chemical and electrical structure of proteins will be enough to understand their conformational changes, and gating; just as a proper understanding of those properties is enough to understand automobile engines, but it is possible, I guess, that conformation changes are so complex, resulting from interactions of so many chaotic systems that they need separate explanation.

What seems impossible is that we need separate laws to describe functions of a protein that occur without conformation change. Vitalism in even its most limited sense has no place in the analysis of function produced by just one protein conformation, and probably not in functions produced by conformation changes either. It is certainly better when planning experiments to assume that conformational changes can be explained by the ordinary laws of physics than to assume otherwise. Indeed, a good way to reveal the role of complexity is to understand what would happen without it. That means trying to make physical models of models of biological systems, using just as much complexity as is needed to explain experiments, but no more, hopefully not using state models whose complexity is comparable to the experimental data set.

Open ionic channels are thought to function mostly in just one conformation, so if there is any biological function that can be described entirely in the language of physics and chemistry, it should be current flow through channels, once they are open. (For present purposes we ignore subconductance states.) That language is also a particularly easy one. Permeation (as ion movement through open channels is called) does not involve changing covalent bonds, and so permeation can be described without quantum chemistry, without traditional organic or biochemistry for that matter. Indeed, if channels are studied only when they are open, when ions are moving through their pores at nearly the rate they move in free solution, it is obvious that channels should be viewed as natural nanotubes through which ions move much as they move in artificial nanotubes, i.e., in crystalline channels (Krager & Ruthven, 1992; Paul, 1982; Perram, 1983; Wilmer et al., 1994) or in free solution (Anderson & Wood, 1973; Berry, Rice & Ross, 1980; Blum, 1975; Blum, Holovko & Protsykevych, 1996; Blum & Hoye, 1977; Bockris & Reddy, 1970; Conway, Bockris & Yaeger, 1983; Fleming & Hänggi, 1993; Friedman, 1962; Friedman, 1985; Hänggi, Talkner & Borokovec, 1990; Henderson, 1992; Hockney & Eastwood, 1981; Murthy & Singer, 1987; Newman, 1991; Tyrrell & Harris, 1984).

Thus, when we seek to understand how the function of channels arises from their structures, the physical laws are very clear. They are the physical laws that govern the behavior of condensed phases like ionic solutions and proteins. Those laws are not numerous. Condensed phases (under these biological conditions) do quite little. Matter in them can diffuse; it can neither fission nor fuse; energy in them can diffuse as well, in the form of heat; both energy and matter can flow according to the laws of convection; and, most importantly, matter and energy can be moved by an electric field.

***Working Hypothesis.*** Here we will consider a simple working hypothesis, and check to see how well it does. We imagine that all permeation properties of open ionic channels can be predicted by understanding electrodiffusion in fixed structures, without invoking conformation changes, or changes in chemical bonds. We know, of course, that ions can bind to specific protein structures, and that this binding is not easily described by the traditional electrostatic equations of physics textbooks, that describe average electric fields, the so called 'mean field'. The question is which specific properties can be explained just by mean field electrostatics and which cannot.

I believe the best way to uncover the specific chemical properties of channels is to invoke them as little as possible, seeking to explain with mean field electrostatics first. Then, when phenomena appear that cannot be described that way, by the mean field alone, we turn to chemically specific explanations, seeking the appropriate tools (of electrochemistry, Langevin, or molecular dynamics, for example) to understand them. In this spirit, we turn now to the structure of open ionic channels, apply the laws of electrodiffusion to them, and see how many of their properties we can predict just that way.

***Structure is geometry and charge***. The structure of an open channel is the location of its atoms, the location of its nuclei and the surrounding electron clouds. Those clouds are more or less directly measured by x-ray crystallography and for our purposes the structure will be the coordinates given us by that technique.

Few ionic channels have been crystallized and 'structured', but with the understanding of just how little can be done without three dimensional structure, effort is increasing and progress forthcoming, see the recent publication (Doyle et al., 1998: structure ***1BL8*** at the Brookhaven web site *http://pdb.pdb.bnl.gov*) of the structure of the $K^+$ channel of *Streptomyces lividans* (that our lab likes to call the *McK* channel, in appreciation for the hard work and significance of the contribution of MacKinnon's lab, as well as the public prominence of the *Mc* prefix.

Clearly, the movement of ions through a channel depends on the geometry of the hole in the protein. Atoms cannot long exist in the same place at the same time, and so a hole of larger diameter will let through more ions in a given time (with a given driving force) than a hole of smaller diameter. A channel of longer length is likely to expose the permeating ion to greater friction, and thus to allow less flow of ions in a given time (with a given driving force) than a shorter channel. The geometry of the open channel is one important feature of the protein.

The reason atoms cannot exist for long in the same place at the same time is sometimes forgotten: nuclei and electrons occupy only a tiny nearly negligible fraction of the space of an atom. The reason that atoms cannot easily overlap is that their electron clouds repel so strongly. The electrical interactions of quantum mechanics determine this mechanical property of atoms.

Similarly, the electrical properties of atoms dominate much of their other behavior, behavior which is often called structural. The charge on atoms is sufficiently large, and the distances sufficiently small that the electrical forces dominate all others. This essential fact, stated so clearly in the first paragraphs of Feynman's magnificent textbook on electricity and magnetism (Feynman, Leighton & Sands, 1963), cannot be reiterated too often, particularly given how widely it is unknown. Electric forces are exceedingly strong, and so must always be considered explicitly when studying channels. They often, in one guise or another, will turn out to dominate the structural properties of systems. The question is how do we describe these forces. Should we use the structural language of mutual exclusion, of simple steric effects, or do we need something more sophisticated, like the language of electrostatics, or quantum chemistry?

The electrical properties of the matter we deal with in channels and proteins are relatively easy to describe. Currents are tiny and magnetism is not involved. Metals do not occur and metallic conduction is rare, nearly non-existent. Proteins and channels have static electric charge, determined by their chemistry, by the solution of Schrödinger's equation. This charge is usually much larger than the induced charge produced by the local electric field, that is by polarization. It is this fixed charge which is the crucial determinant of many properties of proteins. And it is this fixed charge we consider in our working hypothesis: when we say we will consider how well the electrical properties of channel proteins determine the permeation of ions, what we mean is we consider how well the structural charge of proteins determines this permeation.

The existence and size of this structural charge is not emphasized as much as I would wish in elementary textbooks in either electricity and magnetism or molecular biology. The biology textbooks speak at length of polar chemical bonds and polar amino acid residues, but they rarely say polar bonds and residues are simply those with significant (localized) electrical charge. The textbooks of electricity and magnetism pay almost no attention to the boundary conditions of charged matter: they are focussed on the properties of the electric field in vacuum or in dielectric materials that do not contain net fixed charge. And chemistry, physics, and biology textbooks produce considerable confusion by their use of the word polar (meaning permanent distribution of fixed charge independent of the local electric field) and polarization (meaning the induced distribution of fixed charge that is zero when the local electric field is zero).

Education is not helped either by the widespread use of dipoles to describe charge distributions. As appropriate as this description is when studying an electric field far away from a charge distribution (of finite size), it is inappropriate close to the charge distribution, where most of chemistry and nearly all of molecular biology occurs. Close to a 'polar' molecule (like water), which has zero net charge (i.e., the integral of its charge density over all space is zero), the electric field is not even crudely approximated by that produced by a point dipole. Many many terms of a Taylor expansion (called a multipole expansion when it is constructed from Coulomb's law written in polar coordinates) are needed to describe that field, hundreds or thousands of terms might be needed if a permeating ion is nearly touching the fixed charge distribution, i.e., if an ion is hydrated by an adjacent water molecule or solvated by a nearby 'binding site'. The dipole term is just the second term of the multipole expansion and the expansion involves hundreds of terms of nearly equal size, when considering fields close to distributions of fixed charge.

Although these are strong words, they deal with matters of mathematics and the convergence of infinite series, which really are not too ambiguous. Simple computations (e.g., substitution in eq. 3.88-3.91 of Griffiths, 1981, of the multipole expansion for the case of $r = 1.01R$, where *r* is the radial coordinate of the edge of a permeating ion, and *R* is the radial coordinate of the edge of the charge on the channel) will validate my statements about the adequacy of dipole models

***Boundary conditions for proteins***. The boundary condition that describes matter, particularly the electric field produced by the charge of matter, depends on the resolution of the description, and high resolution descriptions undoubtedly would benefit by theoretical analysis beyond that of the mean field. Nonetheless, the fundamental issues are well illustrated by the mean field boundary condition which describes the electric field at the edge of a protein produced by fixed charge at that edge, both being averaged over a long time compared to atomic fluctuations, say averaged over nanoseconds or longer. Then, the boundary condition is

$$\frac{\partial\varphi(\vec{\Gamma}_2)}{\partial n} - \frac{\partial\varphi(\vec{\Gamma}_1)}{\partial n} = -\frac{\upsilon_0\left(\vec{\Gamma}\right)}{\varepsilon_0} - \frac{\sigma_2\left(\vec{\Gamma}_2, \varphi(\vec{\Gamma})\right) - \sigma_1\left(\vec{\Gamma}_1, \varphi(\vec{\Gamma})\right)}{\varepsilon_0} \tag{1}$$

or equivalently, when induced charge is strictly proportional to the local electric field,

$$\varepsilon_{Wall}(\vec{\Gamma})\frac{\partial\varphi(\vec{\Gamma})}{\partial n} - \varepsilon_{Pore}(\vec{\Gamma})\frac{\partial\varphi(\vec{\Gamma})}{\partial n} = -\frac{\sigma_0(\vec{\Gamma})}{\varepsilon_0} \tag{2}$$

Here, $\varphi(\vec{\Gamma})$ is the electric potential on the channel wall, which has a dielectric 'constant' in the range $\varepsilon_{Wall}(\vec{\Gamma}) \cong [10,30]$ compared to the dielectric coefficient $\varepsilon_{Pore} \cong [20,80]$ of the pore. The induced charge $\sigma_2(\vec{\Gamma}_2,\varphi(\vec{\Gamma}))$ is on the channel wall $\vec{\Gamma}_2$ (and depends on the local electric field, of course); the induced charge $\sigma_1(\vec{\Gamma}_1,\varphi(\vec{\Gamma}))$ is located within the pore, just next to the wall, at $\vec{\Gamma}_1$. $\varepsilon_0$ is the permittivity of free space.

The interfacial surface charge $\sigma_0(\vec{\Gamma})$ of these equations is an expression of the covalent bond structure of the protein and the ionization state of the acidic/basic residues. (Note that here we define $\sigma_0(\vec{\Gamma})$ to exclude any component of interfacial surface charge that is proportional to the local electric field. Those components are described by the dielectric constants.) In the context of channels, $\sigma_0(\vec{\Gamma})$ is a permanent structural charge that changes only if the ionization state of the protein changes. That can happen, if the local pH is changed either by changing the pH of the bulk solution, or by changing the electric field enough to change the local concentration of hydrogen ions and the effects can be important. But for the purposes of this review, we will assume that the permanent charge has a fixed value. We have dealt with ionization effects elsewhere (Nonner & Eisenberg, 1998).

The interfacial surface charge $\sigma_0(\vec{\Gamma})$ of these equations is the main source of the electric field in most biological and many chemical systems. This fact is not widely known, unfortunately, and the lack of knowledge has led to significant confusion among biologists, chemists, and biochemists (in particular), in my opinion.

Biochemists and channologists usually (if not invariably) describe the surface of a protein as a potential profile ('potential of mean force') and, forgetting that the potential of mean force is a variable output of the system (Hill, 1956; Hill, 1960; Hill, 1977; Hill, 1985), they treat the potential of mean force as a fixed input or source that does not change with experimental conditions, as if it arose from a Dirichlet boundary condition (the precise name for a boundary condition that specifies the potential) that did not change with experimental conditions. Biochemists and channologists usually (if not invariably) assume that the potential of mean force (or a rate constant derived from that potential, see eq. (5) does not vary when the concentration of ions surrounding the protein are varied (as they often are in experiments). In fact, the electric field arises (mainly) from a boundary condition (i.e., eq. (1) or (2)) which describes the effects of an unchanging charge (when induced charge is negligible, as us often the case in proteins and nearly always the case in channels). If the charge on the surface of a protein does not change with experimental conditions, then the potential on that surface will change when almost any change is made in experimental conditions. Indeed, the potential everywhere (not just at the boundary) will change as experimental conditions change, and that change cannot be expected to be small, nor is it small in the large number of channels and experimental conditions we have studied to date.

Biologists and biochemists are often put off by these discussions of boundary conditions. Boundary conditions sound like mathematical technicalities that are a minor part of a physical problem, particularly if the listener has not taken a course on differential equations. Teachers of physics often inadvertently reinforce this view, because they traditionally emphasize the beauty and generality of the field equations rather than the significance of nitty-gritty boundary conditions.

Whatever the human considerations, it is a simple fact, easily verified by direct computation of the solution of almost any differential equation, that *boundary conditions are usually important, often dominant determinants of the properties of physical systems because they describe the flow of matter, energy, and charge into the system*.

It is obvious in the laboratory that one must control the flow of matter and charge (i.e., the concentration of ions and the flow of current) if one is to do reproducible experiments. Much of our experimental training and apparatus is designed to control this flow and provide reproducible results. We should expect neither more nor less of the theoretical description than experimental reality: if changing the concentration of ions changes an experimental result (e.g., current) as it nearly always does, or if changing the electrical potential changes the current, as it nearly always does, we need to be sure that the variables for concentration and potential are properly described and controlled in any theory seeking to explain those experiments. Certainly, the theory must contain those variables. Attempts have been made to compute the current through a channel using simulations that do not define a *trans*membrane potential or concentration of permeant ions (Roux & Karplus, 1991a; Roux & Karplus, 1991b), and such calculations have in fact been done in large numbers and received substantial support and attention (Roux & Karplus, 1994).

The surface of a protein cannot be described as an unchanging potential for two reasons, which are really restatements of each other. First, the potential changes because the mobile charges near the surface of the protein and on the boundaries change when experimental conditions are changed. The fixed charge of the surface is fixed, but the concentration of mobile ions that are attracted to the fixed charge varies as ion concentrations are varied. That is to say, the shielding of the fixed charge varies with experimental conditions.

The only way the potential could be maintained is if charge were supplied to the surface of the protein, i.e., if the fixed charge were changed. In some systems, charge is in fact supplied in just this way. At a metal electrode connected to a voltage clamp amplifier or battery, charge is supplied. The amplifier or battery supplies the charge to the metal electrode necessary to keep the potential constant. In these systems, matter (i.e., the electrode) cannot be described by a fixed value of charge, but rather by a fixed value of potential. Of course, in this type of system the electrode must be connected to a source to maintain the potential: boundary conditions of this type are sources of energy, etc. Or in mathematical terms, if the potential is independent of experimental conditions, its derivative (which is more or less proportional to flux) cannot be.

The surface of a protein has no access to a source of charge which would be needed to maintain the potential. Unlike the electrode just mentioned, or the wires that conduct electricity in the walls of our buildings, the surface of a protein is not connected to a generator that 'makes' (i.e., separates) charge by the burning of fossil fuel or the fission of uranium. Rather, *the potential at the surface of a protein is determined (mostly) by the shielding of* *the fixed charge on that surface.* The extent of shielding is a sensitive function of ionic conditions and is often the dominant determinant of the electrical properties of ionic solutions and proteins. It is hardly an exaggeration to say that studying shielding has been the central theme of electrochemistry for many years, at least since Debye and Hückel showed that shielding is the dominant determinant of the properties of ionic solutions nearly a century ago. Shielding has been known to be a crucial determinant of the properties of proteins for at least 75 years. See Ch. 5 of (Edsall & Wyman, 1958).

The treatment of the surface of a protein as an unchanging potential of mean force is not compatible with the generally accepted treatment of ionic solutions, and of proteins in solution, because it ignores the effect of experimental conditions, bath concentrations and *trans*membrane potentials, on the potential of mean force. This treatment of a protein is thus not compatible with the Debye-Hückel, Gouy-Chapman, or Poisson-Boltzmann theories (Berry et al., 1980; Bockris & Reddy, 1970; Conway et al., 1983; Davis & McCammon, 1990; Friedman, 1962; Friedman, 1985; Harned & Owen, 1958; Honig & Nichols, 1995; Newman, 1991; Robinson & Stokes, 1959; Schmickler, 1996) or their modern replacement, the Mean-Spherical-Approximation (*MSA*) theories (Bernard & Blum, 1996; Blum, 1975; Blum & Hoye, 1977; Hoye & Blum, 1978).

Of course, the potential of mean force at the surface of a protein can *sometimes* be independent of concentration of reactants, in special circumstances, for example, when the total ionic strength is held constant, while the substrate concentration is not varied enough to itself shield the fixed charge of the other reacting species or protein. Nonetheless, these are special circumstances not likely to be present in most experimental or biological systems, and they are certainly not present in open channels.

***Flux of individual ions***. At first, the difficulties arising from the usual description of the surface of proteins may seem isolated: after all, not many workers or papers are concerned with that subject. However, many, even most workers and papers concerning enzymes, channels, and proteins describe the function of these molecules as chemical reactions, using the 'law of mass action' to describe the function, and that law (as usually used) depends on the description of the surface of a protein. For example, in the case of channels, the chemical reaction

$$L \underset{k_b}{\overset{k_f}{\rightleftarrows}} R \tag{3}$$

is widely used to describe permeation, and of course similar statements are found on nearly every page of a biochemistry textbook. This chemical reaction can be translated, without approximation, into an equivalent statement of flux, as the 'law of mass action'

$$J_k = \overbrace{d \cdot k_f C_k (L)} - \overbrace{d \cdot k_b C_k (R)} \tag{4}$$

where $d$ is the length of the channel, $C_k$ is the concentration of ions on the **L**eft or **R**ight side of the channel, the rate constants $k_f$, $k_b$ have units of $\sec^{-1}$, and $J_k$ is the flux of that ion (units: concentration per cross sectional area per second).

If equation (4) is used as a definition of a rate constant, with the flux being determined independently by other equations, no difficulty arises. But if the rate constants of the chemical reaction are assumed to be independent of concentration, as is nearly always the case, serious problems arise because then the shielding effect of concentrations of ions is not included: as we have already discussed, the flux of ions always depends on the potential, the potential nearly always depends on the concentration of mobile ions, because the shielding of fixed charge depends on that concentration, and so the rate constant must depend on concentration in most cases.

This argument is inescapable, because the 'law of mass action' has no life of its own. It is not an independent physical law, but must be derived from the underlying physical model of the flux and its dependence on structure, mechanism, etc. In the case of channels, this derivation can be made explicit under very general conditions (i.e., the existence of conditional probabilities, Eisenberg, Klosek & Schuss, 1995). The equations are particularly neat when friction is large and simple in behavior, described by a single diffusion coefficient, a single number $D_k$ for each species $k$ of ion.

$$k_f = k\{R|L\} = \frac{D_k}{d^2} \cdot \frac{\exp(z_k V)}{\frac{1}{d}\int_0^d \exp z_k \Phi(\zeta)\, d\zeta} = \frac{D_k}{d^2} \mathrm{Prob}\{R|L\}$$

$$k_b = k\{L|R\} = \frac{D_k}{d^2} \cdot \exp(z_k V) \frac{1}{\frac{1}{d}\int_0^d \exp z_k \Phi(\zeta)\, d\zeta} = \frac{D_k}{d^2} \mathrm{Prob}\{L|R\} \tag{5}$$

The rate constant $k_f = k\{R|L\}$ in fact is (nearly) the conditional probability $\mathrm{Prob}\{R|L\}$ that an ion entering a channel on the left leaves on the right (when absorbing boundary conditions are placed on the right: see the original paper for the precise specification of the probability model and physical system) and that conditional probability can be evaluated either by mathematical analysis (to give the expressions of eq. (5)) or by direct simulation of the motion of individual ions Barcilon *et al.* (1993) computed nearly 2 billion trajectories) fortunately with identical results: compare their eq. 2.24 and 7.5 with eq. 6.15 of Eisenberg et al., 1995. These expressions use normalized units $\Phi(x) = F\varphi(x)/RT$; $V = FV_{appl}/RT$ and can be easily generalized if $D_k$ depends on location (Nonner & Eisenberg, 1998).

The rate constant of equation (5) is very different from the rate constant used in traditional barrier models of ionic channels

$$k_{trad} = kT/h \tag{6}$$

because the Kramers rate constant includes the effect of friction and $k_{trad} = kT/h$ does not. (*h* is Planck's constant more usually found in problems of quantum mechanics.) The implications of this fact are discussed at length in a few pages.

It is important to realize that the description of ionic flux we have just provided is *not* a continuum or macroscopic description of ionic motion (although in fact it can all be described by a diffusion equation, e.g., the Nernst-Planck equation, see (Eisenberg et al., 1995).

$$J_k = -D_k\left(x\right)A\left(x\right)\left(\frac{dC_k\left(x\right)}{dx} + \frac{C_k\left(x\right)}{RT}\frac{d}{dx}\left[z_k e N_A \varphi\left(x\right) + \mu_k^0\right]\right)$$

$$I = \sum_k I_k = \sum_k z_k F J_k \tag{7}$$

The Nernst-Planck equations describe the probability of location of individual atoms, following random trajectories. The flux $J_k$ of ions and the electric current $I_k$ carried by each ion of charge $ez_k$ is driven by the (gradient of) concentration and electrical potential, which together form the chemical potential $\mu_k = RT\log_e C_k\left(x\right) + z_k F\varphi(x)$. $e$ is the charge on a proton, $N_A$ is Avogadro's number, and $z_k$ is the valence of the ion. $\mu_k^0$ is the standard chemical potential that describes energies other than those controlling the electric field and diffusion, e.g., binding or the difference between dehydration and resolvation. The cross sectional area of the channel $A\left(x\right)$ and the diffusion coefficient $D_k\left(x\right)$ can be functions of location. The expressions for the rate constants (5) are in fact solutions of equation (7) as shown in detail in Eisenberg *et al.* (1995).

The averaging, and mean field properties of the models we use arise chiefly in the description of the electric field.

***Poisson's equation.*** It seems inescapable then that we must determine how the rate constant varies with concentration if we are to proceed, and that means determining how the potential profile varies with concentration. In other words, we cannot use just the law of mass action to describe flux, but we must also use Coulomb's law, or its equivalent Poisson's equation, to show how potential (and rate constants) vary with concentration. The theory we use to describe an open channel represents the structure of the channel's pore as a cylinder of variable cross sectional area $A\left(x\right)$ (cm$^2$) along the reaction path $x$ (cm) with dielectric coefficient $\varepsilon_r\left(x\right)$ and a density of charge $\rho\left(x\right)$ (coul cm$^{-1}$). $eN_A$ is the charge in 1 mole of elementary charges $e$, i.e. the charge in a Faraday. The charge $\rho\left(x\right)$ consists of

(1) the charge $eN_A \sum_k z_k C_k(x)$ of the ions (that can diffuse) in the channel, of species *k* of charge $z_k$, and mean concentration $C_k(x)$; typically $k = \mathrm{Na}^+$, $\mathrm{K}^+$, $\mathrm{Ca}^{++}$, or $\mathrm{Cl}^-$ and

(2) the permanent charge of the protein $P(x)$ (mol cm$^{-1}$), which is a permanent part of the atoms of the channel protein (i.e., independent of the strength of the electric field at *x*) and does not depend on the concentration of ions, etc, and so is often called the fixed charge. Permanent charge is really quite large (~0.1−1***e*** per atom) for many of the atoms of a protein. The function *P*(*x*) is a one dimensional representation of the full three dimensional distribution of (fixed) charge in the protein. It includes the integral of the surface charge $\sigma_0(\vec{\boldsymbol{\Gamma}})$ of the protein described in eq. (2). More specifically, and more generally

$$P(x) = -\left[\frac{\varepsilon_r \varepsilon_0}{eN_A}\frac{d^2\bar{\varphi}_{3D}(x)}{dx^2} + \sum_k z_k \bar{C}_{k\,3D}(x)\right] \tag{8}$$

where the dielectric constant $\varepsilon_r$ is assumed independent of location (only for simplicity in writing) and $\bar{\varphi}_{3D}(x)$ represents the cross sectional average of the electrical potential computed from a three dimensional version of *PNP* (Hollerbach et al., 1999). $\bar{C}_{k\,3D}(x)$ is the cross sectional average of the concentration (units: moles/liter) of an ion of type *k* computed from a three dimensional version of *PNP*.

(3) The dielectric charge (i.e., the induced charge which is strictly proportional to the local electric field) is not included in $\rho(x)$ because it is described by $\varepsilon_r(x)$. It is generally very small compared to the structural charge, but might not be in a pore lined with nonpolar residues (see later discussion of the in-pore in the McK channel).

Next, we make the usual mean field assumptions that the average charge $\rho(x)$ produces an average potential $\varphi(x)$ according to Poisson's equation and that the mean electric field $-\nabla\varphi$ captures the properties of the fluctuating electric field which are important on the slow time scale of biology. These assumptions are hardly novel; indeed, it requires some extraordinary circumstances for them not to be true, on the slow highly averaged time scale relevant for ion permeation and most biological processes. *If the* *potential energy of mean electrical force, averaged for 1 msec, did not come from the mean electric charge, which source could it come from?* If there were such a significant force, that did not come from the mean electric charge, or gradients of chemical potential, it would probably have been noticed and given a name, e.g., as binding or flux coupling or some such.

Not wishing to assume such a force, we write Poisson's equation as

$$\varepsilon_0\left[\varepsilon_r(x)\frac{d^2\varphi}{dx^2}+\left(\frac{d\varepsilon_r(x)}{dx}+\varepsilon(x)\frac{d}{dx}\left[\log_e A(x)\right]\right)\frac{d\varphi}{dx}\right]=-\rho(x) \tag{9}$$

where the average charge in the channel's pore is given by

$$\rho(x)\equiv eN_A\left[P(x)+\sum_k z_k C_k(x)\right] \tag{10}$$

$A(x)$ describes the cross sectional area of the pore at location *x,* $\varepsilon_r(x)$ is the dielectric constant (relative permittivity) at location *x*, and $\varepsilon_0$ is the permittivity of free space. The small dielectric term is neglected.

The boundary conditions for the potential in the real world are set by the experimental conditions: it has been known since the time of Hodgkin and Huxley (Hodgkin, Huxley & Katz, 1949) (Cole, 1947) that experiments are most easily interpreted if done under 'voltage clamp' conditions, so complex uncontrolled effects of voltage are avoided. Special apparatus is used to control the potentials in the baths surrounding the channel, i.e., the potential on the left is known and maintained at $V_{appl}$ and that on the right is held at zero.

$$\begin{aligned}\varphi(L)&=\varphi(-\infty)=V_{applied}\\ \varphi(R)&=\varphi(+\infty)=0\end{aligned} \tag{11}$$

These boundary conditions are maintained by charge supplied to the system at its boundaries (i.e., by electrodes placed in the bath and/or inside a cell or pipette). The amount of charge necessary to maintain the potentials depends on the properties of the system, e.g., of the channels, and the experiment (i.e., whether solutions or *trans*membrane potential $V_{applied}$ are changed). This is the charge supplied by the voltage clamp apparatus used in measurements of ionic currents.

Of course, the natural activity of membranes and channels does not occur when the voltage clamp apparatus is used. Nonetheless, natural voltage changes can easily be reconstructed by solving the Hodgkin-Huxley equations (Hodgkin & Huxley, 1952), which show how the current through a (voltage clamped) membrane produces the uncontrolled transmembrane potentials of a normally functioning cell. Weiss (Weiss, 1996) is a nice description of the classical biophysics and physiology which arose from the work of Hodgkin, Huxley, and Katz, more than anyone else. All modern systems for studying the current through one channel protein use the voltage clamp, e.g., the "patch clamp" of Sakmann and Neher (Sakmann & Neher, 1995), see also Levis & Rae (1992,1995).

The concentrations of ions must also be controlled if the properties of channels are to be easily understood, implying the boundary conditions

$$C_k\left(L\right)=C_k\left(-\infty\right), \qquad C_k\left(R\right)=C_k\left(+\infty\right) \tag{12}$$

Special apparatus is not available to maintain this boundary condition, but the large volume of the baths surrounding channels, and the relatively small amounts of charge transferred through a single channel (in many cases) often guarantees that concentration changes produced by flux are not significant. Such is *not* always the case, indeed such may never be the case for $Ca^{++}$ channels functioning in their normal mode, and certainly the absence of noticeable concentration changes must always be verified experimentally for any channel. Nonetheless, the checks are easily done and usually satisfied.

These boundary conditions (11) and (12) (at $x=\pm\infty$ of the three dimensional problem), do not map obviously and easily into boundary conditions at the ends of the channel $x=0,\ x=d.$ We have used a particular well–precedented equilibrium mapping called the built-in potential in semiconductor physics or the Donnan potential in parts of biology (see Barcilon, 1992; Barcilon, Chen & Eisenberg, 1992; Chen, Barcilon & Eisenberg, 1992; Chen & Eisenberg, 1993a). Other treatments of the ends of the channel are used in the later versions of *PNP* (Nonner, Chen & Eisenberg, 1998; Nonner & Eisenberg, 1998) and three dimensional versions of the model have been constructed that have no arbitrary boundary conditions (Hollerbach et al., 1999). Preliminary results suggest that the simple boundary conditions yield surprisingly adequate representations of the current and spatially averaged properties of the channel, although of course any averaged treatment misses *atomic* details of considerable interest and importance.

***Coupling and solving: the Gummel iteration***. The electrical potential is described by Poisson's equation, as we have seen; and the flux is described by the Nernst-Planck equations, or by rate constants (which are precisely equivalent if defined as in eq. (5)). But neither equation can be solved by itself. The concentrations of ions that flow in the Nernst-Planck equation are the same concentrations of charge that produce the electric field in the Poisson equation, and the electric field of the Poisson equation modifies the flow. The equations are coupled, and must be solved together.

The Gummel iteration (Gummel, 1964; Scharfetter & Gummel, 1969) was discovered decades ago by the semiconductor community (Bank et al., 1990; Bank, Rose & Fichtner, 1983; Hess, 1991; Hess, Leburton & Ravaioli, 1991; Jerome, 1995; Kerkhoven, 1988; Kerkhoven & Jerome, 1990; Kerkhoven & Saad, 1992; Lundstrom, 1992) and was discovered in my lab independently by Duan Chen, some years later (e.g., Chen & Eisenberg, 1993a). The iteration is a general method for producing a self-consistent solution of coupled equations closely related to the self-consistent field methods used in quantum mechanics to compute orbitals. It is described at some length in our publications (*loc. cit.*) and code implementing it is available on our ftp site ftp.rush.edu in directory /pub/Eisenberg.

***Comparison with experiments.*** The *PNP* equations form a map between the structure of the channel protein, represented crudely by the function $P(x)$ and the current voltage curves measured experimentally.

Different types of channels have different pores made with linings of different charge. A useful and productive working hypothesis assumes that the only difference between different types of open channels is their different distributions of fixed charge $P_i(x)$, as defined in eq.(8), where the subscript $i$ identifies the type of channel protein, e.g., a voltage activated $Na^+$ channel, a stretch activated channel and so on (Conley, 1996a; Conley, 1996b; Conley, 1997; Peracchia, 1994; Schultz et al., 1996). Of course, this working hypothesis cannot always be true: specific chemical interactions, not captured in this simple mean field theory, will no doubt be important in ways we do not yet understand. Nonetheless, as we write these words, the current voltage relations of some 7 types of channels in a wide range of solutions can be predicted by simple distributions of fixed charge $P_i(x)$ (Chen et al., 1998a; Chen et al., 1997a; Chen, Lear & Eisenberg, 1997b; Chen, Nonner & Eisenberg, 1995; Nonner et al., 1998; Nonner & Eisenberg, 1998; Tang et al., 1997). The data from the porin channels is of particular interest because the locations of the atoms of that protein are known by x-ray crystallography (Cowan et al., 1992; Jeanteur et al., 1994; Schirmer et al., 1995) and the analysis using *PNP* recovers the correct value of charge when a mutation is made in the protein.

One particular kind of channel (the calcium release channel *CRC* from cardiac muscle) has been the object of extensive experimentation. This channel also appears to be strikingly simple: a fixed charge $P_{cardiac}\left(x\right)=P_0$ independent of position, with $P_0$ equal to ~1***e***, predicts the currents measured in solutions containing a single species of each of the monovalent cations (i.e., $Li^+$, $Na^+$, $K^+$, $Rb^+$, $Cs^+$, as the chloride salt) from 20 mM to 2 M concentration, and potentials in the range ±150 mV, assuming each ion has a different diffusion coefficient (Chen et al., 1998b). The value of the diffusion coefficients inside the channel are estimated by fitting theoretical predictions to the experimental data. Typically, the diffusion coefficients inside the channel found to be some 10× less than in free solution. The $Li^+$ data is not fit as well as the other ions', but a small change in the theory, required in any case to fit data in mixed solutions, improves the fit significantly, as described later in this paper.

This result surprised us considerably, because it shows that the same permanent charge and structural parameters (e.g., diameter and length) can fit an enormous range of data, implying that the channel is much the same whether an ion with a diameter of around 1.4Å ($Li^+$ ) or 3.9Å ($Cs^+$) fills the channel's pore. Of course, that is something of an overstatement, since the value of the diffusion coefficient inside the channel is different for each ion and can be determined only by estimation from the experimental data. But the value of the diffusion coefficient for an ion is the same in all solutions, no matter what their concentration or composition, and at all potentials, and so the naïve interpretation seems safe to me: the *CRC* channel is much more rigid than any of us have expected (as measured by the average value of the properties that determine flux on the biological time scale). The data seems to show that all monovalent ions interact with the same mean electric field, which does not depend on the diameter or chemical nature of the permeating ion. I hasten to add, however, that this result, while clearly true for the *CRC* channel may not be true for other channel types.

***Selectivity: properties in mixtures of ions.*** The experiments just described were performed in homogeneous solutions of the different types of ions, e.g., 20 mM NaCl on one side of the channel with 200 mM $Na^+$ on the other, or 50mM CsCl on one side and 500 mM CsCl on the other. A more common (but complex) way to study selectivity is to make mixtures of ions and apply them to both sides of the channel, e.g., 20 mM NaCl and 20 mM CsCl on one side and 200 mM NaCl and 200 mM CsCl on the other. The ability of channels to select between ions is one of their most important and characteristic properties and so experiments of this type have received much attention, with probably hundreds of papers being written in the last few years on the different selectivity of different channels under varying conditions.

Before we consider to the properties of channels in such mixtures, it seems sensible (following Chen, 1997) to examine the properties of mixed solutions in the bulk (Anderson & Wood, 1973; Robinson & Stokes, 1959), i.e., in the absence of channels. Those properties are much more complex than imagined in most channel texts, particularly when concentrations are large. Since 1 ion in a region $7 \times 10$ Å is a concentration of around 5 M, ions in channels must be expected to resemble ions in highly concentrated (nearly saturated) solutions, not ions in highly dilute solutions. In highly concentrated bulk solutions, the movement of ions is highly correlated, linked by the electric field and does not resemble independent movement at all.

The image of ions moving independently in ionic solutions (or in channels for that matter) can only be true when they are so far apart that their electric fields do not interact; this image is not true even in the very dilute solutions which can be adequately described by the Debye-Hückel/Gouy-Chapman/Poisson-Boltzmann theories, because the essence of these theories is electrostatic interaction, i.e., shielding. That is to say, ionic concentrations have to be much lower than micromolar for the image of independent ionic movement to have any validity, if it has any validity at all.

The properties of highly concentrated ionic solutions and mixtures in bulk solution can be quite complex and yet can be well described by a remarkable modern theory, called the *MSA* (mean spherical approximation) developed by many workers over the last few decades, but by Lesser Blum, more than anyone else (Bernard & Blum, 1996; Blum, 1975; Blum et al., 1996; Blum & Hoye, 1977; Durand-Vidal et al., 1996; Hoye & Blum, 1978). This is not the place, nor am I the person to review this theory. Suffice it to say that by describing the packing of spherical ions correctly, and the consequent effect of the excluded volume directly on the free energy, and separately on the electric field, the *MSA* is able to predict the activity of ionic solutions from infinite dilution to saturation, even when saturation occurs at many molar! The properties of these solutions are very different from the properties of particles moving independently that pervades the traditional physiological literature.

The case of the *CRC* channel we have already discussed (when bathed in homogeneous solutions) is particularly striking. Here, *PNP* has been used in a wide range of mixed solutions (Chen, *et al.* 1998a). The theory must be slightly modified to accommodate mixtures: the smaller ions ($Li^+$ and perhaps $Na^+$, the data is not clear in the latter case) have an excess free energy beyond that computed from the Poisson equation. Some 1-2 $kT$ of energy (i.e., $\mu_{Li}^0$) must be added to the electrical energy for $Li^+$ to account for the experimental data, but remarkably this number is a constant that does not change significantly over the whole range of conditions examined experimentally, in a range of mixtures of ions. (Everything is not perfect, of course, this being biology, and scientists being human. In one asymmetric solution, there is a systematic misfit we do not understand, and thus call a conformation change. In other solutions, there are small but reproducible misfits. But investigation of these in the absence of a three dimensional structure seems not a useful exercise.)

The existence of an extra energy (i.e., $\mu_k^0$) is hardly a surprise; it is this type of energy that is needed to explain the selectivity properties of highly concentrated bulk solutions. In channels, additional chemical energies is present beyond those in bulk solution: the process of dehydration from bulk solution, resolvation by channel protein and channel water, which accompanies the movement of any ion into (or out of) a channel involves energies 50 to 100× larger (i.e., some hundred $kT$) than the excess energy we find to be present in *CRC*. Indeed, we have been expecting to find signs of such phase boundary 'potentials' (i.e., energies) since long before we wrote the *PNP* equations. What is striking is not the existence of such excess free energy, but rather how little is needed (to fit a wide range of experimental data) and how simple its properties seem to be. Simulations (Dieckmann et al., 1999) suggest that dehydration/resolvation energies are 2 $kT$ or less, a result that is in welcome support of our curve fitting. It seems that the mean field electric forces described by *PNP* dominate the properties of the open channel, even when other forces are present.

The reasons for the dominance of the electric field are not known for certain, and the role of the atomic interactions traditionally thought to be so important in ionic channels (i.e., single filing phenomena, ion-ion repulsion, etc) are not known either. Both issues are important and need investigation. What is known is that in closely related, but not identical systems, physical chemists and physicists have already shown that mean field terms dominate. For example, Henderson, Blum and co-workers (Blum, 1994; Bratko, Henderson & Blum, 1991; Henderson, Blum & Lebowitz, 1979) show that when fixed charge densities are large, as they are in channels, the mean field dominates the properties of systems in a variety of geometries, e.g., the planar geometry of lipid bilayers analyzed in (for example) Gouy-Chapman theory. Indeed, when fixed charges (and the accompanying concentration of counter ions) are 0.5 M, which is one-tenth of the value likely to be present in ionic channels, the mean field is strong enough to swamp ion-ion interactions other than mean field. Experimental evidence (Ben-Tal et al., 1996) shows clearly that mean field theories (Gouy-Chapman) work in the biological domain in planar systems, as predicted by theory. Of course, narrow single file channels are not planar systems; their geometry enforces correlations different from those in planar systems, but these geometrical properties of channels have been reported to make the mean field more (*not* less) dominant in the other systems studied up to now (van den Brink & Sawatzky, 1998).

Clearly, analyses must be done for structures reminiscent of channels before they can be fully convincing. Nonetheless, it seems likely that the high charge density and nearly one dimensional geometry of biological channels are what make most of their properties predictable by a mean field theory like *PNP*, even in the face of single filing. The recent paper of Nelson and Auerbach (1999) seems very important in this context since it is apparent the first to analyze and simulate single file systems of finite length. Nelson and Auerbach show that particle displacements fall into three domains depending on the time scale. A short time domain, in which diffusion occurs much as it does in free solution; an intermediate time scale, comparable to the first passage time of a particle across the channel, in which diffusion behaves much as it does in an infinitely long single file system; finally, a long time domain, in which diffusion occurs much as it does in free solution, but with an apparent diffusion coefficient much less than that in free solution (or in the short time domain, just mentioned). It seems clear that both the measurement and function of biological channels falls into the long time domain. This work would be definitive, in my view, if were extended to analyze ratios of unidirectional fluxes, and the properties of charged particles, moving in the presence of a gradient of electrical potential.

***Anomalous Mole Fraction Effect***. The main signature of single file behavior in single ionic channels is called the anomalous mole fraction effect *AMFE* (Eisenman, Latorre & Miller, 1986), also known as the mixed alkali effect in synthetic crystalline channels (Wilmer et al., 1994). The *AMFE* can be easily be explained by the *PNP* model if a bit of localized chemical binding is introduced (Chen, 1997; Nonner et al., 1998).

Interestingly, the mechanism by which the *AMFE* arises in a *PNP* system is novel, not proposed previously as far as I know. The *AMFE* arises in a way that depends entirely on the properties of the Poisson equation: the binding region of the channel accumulates charge. That charge repels all nearby mobile charges of similar sign. The repulsion creates a depletion layer in series with the binding region which has few ions and thus high resistance. Conduction is determined by the region of high resistance even though it is spatially small. The binding region decreases conductance in this indirect way, *not by decreasing the diffusion coefficient or mobility*. (To keep things simple, in this calculation, the diffusion coefficient of the bound ion is the same as it is in bulk solution and everywhere else.) The repulsion that creates the depletion layer would not occur in a system forced to be electrically neutral or forced to have a prescribed electric field (i.e., a system that did not follow Poisson's equation).

The depletion layer is important because it provides an obvious way that a spatially localized property of a protein under easy genetic control (e.g., the charge on a particular residue of a protein) can dominate conduction through the pore. Changes in the size of the depletion layer can easily modulate or gate the conductance of a channel in a protein just as they modulate and gate the conductance of a channel in a Field Effect Transistor (which obeys quite similar equations). Indeed, in transistors, which are three terminal devices, variations in the depletion layer allow amplification of currents (i.e., flux coupling). It will be interesting to see if this mechanism is actually used by mediated transporters, which may prove to be three terminal devices (Chen & Eisenberg, 1992; Eisenberg, 1996a). Perhaps voltage gated channels should be viewed as three terminal devices, with voltage sensor of traditional electrophysiology (Hille, 1992) being the gate controlling flow through the channel, the gating charge of traditional electrophysiology being analogous to the nonlinear capacitive charge necessary to change the potential on the gate of a field effect transistor. The steep voltage dependence of voltage dependent channels would then arise from the transconductance that allows a small movement of (capacitive) charge on the gate to control a large flow of (ionic) current through the channel.

***L-type calcium channels***. The *AMFE* of L-type calcium channels has received a great deal of attention (e.g., Almers & McCleskey, 1984; Almers, Palade & McCleskey, 1984; Armstrong & Neyton, 1992; Chen, Bezprozvanny & Tsien, 1996; Dang & McCleskey, 1998; Heinmann et al., 1992; Hess, Lansman & Tsien, 1986; Hess & Tsien, 1984; Lee & Tsien, 1983; Tsien et al., 1987). Indeed, it is probably not an exaggeration to say that the properties of these channels have formed the paradigm (Almers & McCleskey, 1984; Hess & Tsien, 1984; Hille, 1975; Hille, 1992) taught to most students of permeation for nearly twenty years. Thus, it is important to see whether *PNP* can account for this data.

The properties of calcium channels are quite complex and so are discussed in detail elsewhere (Catacuzzeno et al., 1999a; Nonner & Eisenberg, 1998). Blocking has not been addressed yet with a self-consistent theory, because that requires a time dependent selfconsistent theory not yet available, although the underlying stochastics have been examined in cases where the potential profile has been assumed, and not calculated from an underlying distribution of charge (Barkai, Eisenberg & Schuss, 1996).

When reading the literature of calcium channels, it is important to realize that the data on the L-type calcium channel (Almers & McCleskey, 1984; Hess & Tsien, 1984) do not establish the existence of an *AMFE* in conductance, but rather describe a complex concentration dependence of *current,* that might be called a mole fraction effect *MFE*, to distinguish it from the *AMFE* of conductance. The distinction between current and conductance is not purely semantic; it has been as central to channology since 1952 (Hodgkin & Huxley, 1952) as it has been to the physics of electricity since 1826 when Ohm introduced the idea (according to p. 90 of Whittaker, 1951).

Conductance is a much more direct measure of the properties (e.g., mobility) of ions in a channel than is current, since current depends on many other variables besides mobility, e.g. voltage. It is not surprising then that the *MFE* of current found in L-type calcium channels, bathed in mixtures of $Ca^{++}$ and $Na^{+}$, is much easier to explain than the *AMFE* of conductance found in $K^{+}$ channels. The *MFE* of calcium channels (as viewed by *PNP*) does not involve a depletion layer but is a consequence of spatially uniform fixed charge (Nonner & Eisenberg, 1998). Given the importance both logically and historically of the *AMFE* in calcium channels, it is surprising that more experiments along the lines of (Eisenman et al., 1986; Friel & Tsien, 1989), have not been performed measuring the *I-V* relations of these channels in a wide range of solutions, seeking conditions in which an *AMFE* is present as well as an *MFE*.

The *MFE* effect is easily explained in a selfconsistent model of calcium channels. Nonner & Eisenberg (1998) modifies *PNP* (into *PNP2*) by including binding of calcium and sodium as an excess chemical potential $\mu_{Ca}^{0}$ , as first suggested by Chen, 1997 . The excess chemical potential of calcium might arise from dehydration of the ion (from the water of bulk solution) and resolvation (by the channel protein and channel water) or from effects of the finite volume of the ions (Bernard & Blum, 1996; Blum, 1975; Blum, 1994; Blum et al., 1996; Blum & Hoye, 1977; Durand-Vidal et al., 1996; Hoye & Blum, 1978) as described in the *M*ean *S*pherical *A*pproximation (*MSA*) of physical chemistry. $\mu_{Ca}^{0}$ is described in *PNP2* by a single number, at all concentrations, at all potentials and in all solutions as long as the pH does not change. A binding of some 3-4 *kT* for calcium, and a repulsion of 2-3 *kT* for sodium are enough to predict the *MFE* found in calcium channels (see Fig. 5 & 6 of Nonner & Eisenberg, 1998). The binding/repulsion is supposed to arise from the glutamates of the channel and the pH dependence of the channel expresses the variable ionization (i.e., a fixed charge that changes with pH) resulting from both the $pK_a$ of their carboxyls (in bulk) and the local electrical potential energy. We are currently trying to show how the excess chemical potential of channels can be explained by excluded volume effects, using the *MSA* (Catacuzzeno et al., 1999a; Catacuzzeno, Nonner & Eisenberg, 1999b).

Another useful approach may be the density functional theory (*DFT*) of heterogeneous systems, e.g., channels in membranes in ionic solutions (Henderson, 1992). Asymptotic analysis (Blum, 1994; Bratko et al., 1991; Henderson et al., 1979) shows that the high charge density lining channels will have a dramatic simplifying effect on the theory, as will the nearly one dimensional distribution of charge (van den Brink & Sawatzky, 1998). Frink & Salinger (1999) has shown that full numerical analysis using *DFT* is feasible, at least in the largest computers available today.

***Traditional explanations for the MFE***. Traditional explanations of the *MFE* of calcium channels are examined in the Appendix of Nonner & Eisenberg, 1998. Traditional models suffer from two significant problems. They ignore friction and they miscalculate the electric field.

Consider the electric field. Traditional models of the *MFE* (Almers & McCleskey, 1984; Almers et al., 1984; Armstrong & Neyton, 1992; Heinmann et al., 1992; Hess et al., 1986; Hess & Tsien, 1984; Lee & Tsien, 1983; Tsien et al., 1987) use *ad hoc* repulsion factors to describe the electrostatic interaction of ions and constant field theory to describe the interactions of ions with the *trans*membrane potential. Both are clearly incorrect.

The electric field along a channel cannot be constant either in space or in experiments, as conditions change (Eisenberg, 1998b; Eisenberg, 1996a; Eisenberg, 1996b;

Syganow & von Kitzing, 1999). A constant electric field can occur only if the lining of the channel's pore is connected to a source of energy and charge and that is clearly not the case. Or, to put the same thing another way, the lining of the channel's wall is a region of fixed charge, not a region of maintained potential. However justified by history (Goldman, 1943; Hille, 1992; Hodgkin & Katz, 1949), the constant field approximation must be replaced because it mistakes the essential property of the electric field in channels, namely that the electric field varies in experiments and space. The variation of the electric field contributes importantly to the biological functions of channels and ignoring that variation makes those functions hard to understand.

Traditional models of the *MFE* have another difficulty in their treatment of the electric field. Traditional theories use arbitrary repulsion factors to describe electrostatic interactions in a way not used by physical scientists for many years. Indeed, the absence of a permittivity of any form in the repulsion factors of traditional theories shows that traditional theories ignore electrical interactions altogether. Coulomb's law (whether written as an integral or in the differential form called Poisson's equation) has been the customary formulation used to describe the repulsive (or attractive) forces produced by electric charge for 173 years (p. 57 of Heilbron, 1979; Whittaker, 1951). Using other treatments of repulsion implies the existence of forces not described by Coulomb's law, i.e., nonelectrical forces.

Such novel forces may exist, of course; for example, effects of the finite volume of ions create forces, in effect; but postulating new physical forces is not the first step one should take in analyzing experiments on channels, at least in my opinion. Few physical scientists would justify the invocation of new physical forces at all; none would justify the invocation of a physical force that has no origin and that follows no specific general rule. If such forces are postulated, it should be at the end of a long line of investigation, and (of course) the forces should be described in such a way that they can be sought in physical systems better defined and more easily studied than open ionic channels.

Barrier models deal even worse with friction than they do with electrostatic repulsion. They ignore friction altogether even though ions move through channels in a condensed phase containing (almost) no empty space. Nothing can move in a condensed phase like a channel's pore without collision and friction. Friction is an invariable concomitant of flux in any condensed phase, and friction is particularly important on small length scales such as in channels (Berg, 1983; Purcell, 1977).

Traditional barrier models do not contain friction, either as a phenomenon or as a parameter, as was pointed out in this journal some 11 years ago (Cooper, Gates & Eisenberg, 1988a). I hasten to add that we were certainly neither alone nor the first to realize the significance of this problem. In the biological literature, see Andersen & Koeppe, 1992; Barcilon et al., 1993; Chiu & Jakobsson, 1989; Cooper, Jakobsson & Wolynes, 1985; Cooper et al., 1988a; Cooper, Gates & Eisenberg, 1988b; Crouzy, Woolf & Roux, 1994; Eisenberg et al., 1995; Läuger, 1991; Roux & Karplus, 1991a). In the chemical literature, the appropriate form for barrier theory in the presence of friction has been known for more than 50 years in chemistry as part of the diffusion theory of chemical reactions (Berne, Borkovec & Straub, 1988; Chandler, 1978; Cho et al., 1993; Coffey, Kalmykov & Wladron, 1996; Dresden, 1987; Eisenberg et al., 1995; Fleming & Hänggi, 1993; Fleming, Courtney & Balk, 1986; Friedman, 1985; Gardiner, 1985; Haar, 1998; Han, Lapointe & Lukens, 1993; Hänggi et al., 1990; Hynes, 1985; Hynes, 1986; Kramers, 1940; Laidler & King, 1983; Murthy & Singer, 1987; Nitzan & Schuss, 1993; Pollak, 1993; Pollak, 1996; Risken, 1984; Tyrrell & Harris, 1984).

***Barrier Models***. It is natural to wonder whether discussion of traditional barrier models is still necessary. After all, incorrect theories often take a generation to be replaced, but science eventually moves on. The reason I write at such length is that the stakes here are large enough to justify the effort, in my opinion. If the tools of physical science are applied to the molecules of biology, it may be possible to create a biotechnology of channels as extensive and efficient as semiconductor technology, but one that operates directly on ions in solution. This technology might allow the manipulation of ions with the complexity and control by which integrated circuits control electrons; yet it would use substrates (ions in water) directly relevant life. The medical and economic consequences of such a technology are obvious.

Technology of this sort depends on an underlying theoretical understanding: integrated circuits would not be possible if the drift diffusion equations (that we call *PNP*) were an inaccurate model.

Until the old verbal models of traditional biology are replaced with physical models, theoretical understanding will be impossible and the technology of channels will be hard to develop. The great majority of workers continue to use barrier models, despite eleven years of criticism. So it seems that we must continue to try to inform biologists of the appropriate form of barrier models which comes from the diffusion theory of chemical reactions.

***Brief History of Diffusion Theory of Chemical Reactions***. The theory of chemical reactions as diffusion of reactants over an energy barrier has been found in textbooks of chemical kinetic for some time (Berry et al., 1980; Coffey et al., 1996; Eu, 1992; Friedman, 1985; Gardiner, 1985; Han et al., 1993; Risken, 1984; Schuss, 1980b; Steinfeld, Francisco & Hase, 1989a; Tyrrell & Harris, 1984; van Kampen, 1981). These theories have several names. Eyring rate theory, transition state theory, activated complex theory, are names found in the chemistry literature. I call them barrier theories here following the channology convention.

The diffusion theory of chemical reactions was introduced, as far as I know, to the biophysical/channel literature by Kim Cooper, then a graduate student of the biophysicist Eric Jakobsson and physical chemist Peter Wolynes (Andersen & Koeppe, 1992; Barcilon et al., 1993; Chiu & Jakobsson, 1989; Cooper et al., 1985; Cooper et al., 1988a; Cooper et al., 1988b; Crouzy et al., 1994; Eisenberg et al., 1995; Läuger, 1991; Roux & Karplus, 1991a). Wolynes (Skinner & Wolynes, 1978; Wolynes, 1980) had an important role in popularizing and extending Kramers' approach to chemical reactions.

Diffusion theory of chemical reactions was more or less started by Kramers (Coffey et al., 1996; Dresden, 1987; Haar, 1998; Kramers, 1940; Laidler & King, 1983). Since then, the diffusion theory of chemical reactions has been one of the pillars of physical and theoretical chemistry. In the last 58 years, some 700 papers have re-derived, extended, simulated, and experimentally tested Kramers' original description of chemical reactions (Fleming & Hänggi, 1993; Hänggi et al., 1990).

There is no controversy in the chemical literature about Kramers' work. Exactly the same results are found throughout the literature of experiments, theory, and simulation, whether the authors come from the Kramers' tradition of diffusion theory (e.g., Fleming & Hänggi, 1993; Hänggi et al., 1990) or from the Eyring tradition of equilibrium statistical mechanics (e.g., Berne et al., 1988; Chandler, 1978; Hynes, 1985; Hynes, 1986; Johnson, Eyring & Stover, 1974; Laidler & King, 1983; Levine & Bernstein, 1987; Pechukas, 1976; Robinson & Holbrook, 1972; Steinfeld, Francisco & Hase, 1989b). These traditions have in fact been united in an elegant and rigorous manner by Schuss, Pollak and co-workers (Pollak, 1993; Pollak, 1996; Pollak, Berezhkovskii & Schuss, 1994). The difficulties about recrossings that concerned Frauenfelder and his colleagues (Fleming & Wolynes, 1990; Frauenfelder, Sligar & Wolynes, 1991; Frauenfelder & Wolynes, 1985)—following Eyring, (e.g., Wynne-Jones & Eyring, 1935)—have been resolved now that the transmission factor has been evaluated by purely mathematical means (Klosek, Matkowsky & Schuss, 1991; Pollak et al., 1994; Schuss, 1980a; Schuss, ) in the case relevant for us (high friction). Chemical reactions in one dimension (i.e., diffusion over a one dimensional barrier) can be considered a closed subject when friction is simple enough to be characterized by a single number, the diffusion coefficient. Applications of chemical kinetics to channels are made easier by the recent finding of simple analytical expressions for the flux *over potential barriers of any shape or height* (Eisenberg et al., 1995).

The description of chemical reactions in high dimensional phase space, when friction is complex, is certainly *not* a closed subject (see, for example, Berne et al., 1988; Eu, 1992; Fleming & Hänggi, 1993; Hynes, 1985; Hynes, 1986; Keizer, 1987). Indeed, quite refined diffusion theories cannot capture the realistic detail typical of even a simple chemical reaction, hydration of $Na^+$ (Rey & Hynes, 1996). Fortunately, permeation through channels is likely to be described well by a one dimensional model with simple friction, because channels are so narrow, and the biological time scale is so slow. Recent experimental work shows that a one dimensional theory with simple friction is surprisingly able to describe many aspects of permeation and selectivity (Chen et al., 1998a; Chen et al., 1998b; Chen et al., 1997a; Chen et al., 1997b; Chen et al., 1995; Nonner et al., 1998; Tang et al., 1997). It may turn out that channel permeation is better described as a chemical reaction than most functions of enzymes (Eisenberg, 1990) because the reaction coordinate of enzymes occurs in a high dimensional phase space, and thus can be tortuous or even ill-defined, whereas the reaction coordinate of ion movement in a channel is simply a line (Elber et al., 1995).

If ions moving through a channel cross a large barrier, the Kramers expression can easily be used. It is barely more complicated or difficult than the traditional expression of channology. Indeed, the more general expression for ionic motion over a barrier of arbitrary shape (not just the high barrier of Kramers theory) is quite simple (eq. (5) above) and can be computed almost as easily as the Kramers expression using Gaussian quadrature formulas. There seems to be no justification for using the traditional barrier expression (6); it offers no significant simplification and the errors involved are enormous, a factor of some 20,000 (Chen et al., 1997a; Cooper et al., 1988a)

***How could such a wrong theory continue to be used?*** It might seem strange that a barrier theory with such a large error would survive. How could a theory in error by a factor of some 20,000 fit the data at all? The logical answer is clear. The error produced by ignoring friction was more or less compensated by the error in assuming, instead of computing the electric field. Together both errors allowed the prediction of a current of the right order of magnitude. In my opinion, barrier theory continues to be used other reasons, more sociological and psychological, than logical. It is difficult for biologists to change paradigms, when they do not understand the physics underlying the original paradigm or its replacement.

On a more practical level, it is difficult to replace the traditional barrier expression with the Kramers expression because barrier models do not come close to fitting experimental data once the Kramers' expression is used. Currents in that case can hardly exceed 0.1 pA (see Appendix of Nonner & Eisenberg, 1998). Few channels have been found that conduct this little current (although many may exist) because such small currents are hard to measure in the background instrumentation noise of patch clamp amplifiers that we use today (Levis & Rae, 1992; Levis & Rae, 1995; Rae & Levis, 1992).

Thus, the experimentalist has faced a dilemma. He (or she) cannot use the correct version of barrier theory to analyze data because it does not fit. Until a theory that fit the data was available, he could either use an incorrect theory, or abandon quantitative analysis altogether, adopting verbal models of permeation.

***Verbal Models in Molecular Biology***. While to physical scientists, verbal models (e.g., of molecules) are superficial popularizations not worthy of professional attention or discussion, verbal models of molecules are used widely in channels and proteins, nearly to the exclusion of quantitative treatments. Sadly, there are distinguished papers, of great quality and importance (e.g., see Doyle et al., 1998) that include such discussions at length. The wide spread use of verbal models in molecular biology and channology forces me to discuss them explicitly here, hoping to discourage their future use, although I am fully aware that physical scientists will be bored by the following words, while molecular biologists will offended, hearing the words as invective, rather than as the analysis that they are intended to be.

Molecular biologists prefer verbal models because most biologists are untrained in applied mathematics and so are unable to deal with quantitative models. This is hardly surprising. The magnificent success of molecular techniques requires much training and hard work and it is rare that any one person can fulfill the demands of molecular biology at all, let alone with time to spare to study physical sciences and applied mathematics. The study of those quantitative sciences takes time and training (and aptitude) just as does the study of molecular biology. And the effort involved in the study of mathematical and physical science is considerable, particularly given their long history and large literature.

But as difficult as the quantitative sciences may be, they must be used if ionic channels or proteins are to be understood, even qualitatively. Structure is a set of numbers specifying the density of electrons (when determined by x-ray crystallography) and is measured in units of centimeters. Permeation is a set of numbers specifying how current varies with potential and concentration. Permeation is measured in units of amps. Words cannot measure densities nor can they compute currents. Numbers and equations are needed for that. Words are simply unable to describe channel structures and permeation with sufficient objectivity and precision to allow unique predictions or scientific testing of alternative models.

***The Necessity of Numbers***. Numbers are needed to understand qualitative properties as well as quantitative properties of channels. If a property (e.g., the current through a channel) is determined by two effects, by the difference of two terms, or the ratio of two factors, then qualitative understanding of the two effects is not enough. The effects may act in opposite directions, and each effect is likely to change in response to some experimental manipulation. The *qualitative* properties of the system, and the nature of its response to the experimental intervention, is determined by the relative size of the effects. For this reason, ***predicting the qualitative function of open channels requires a quantitative theory***. Words cannot evaluate the size of effects. Numbers evaluate the size of effects.

Barrier models illustrate these generalities and they show how easily verbal models can be distorted so they more or less have to fit data. The barrier model of Hille (Hille, 1992; Hille & Schwartz, 1978) has often been modified in an arbitrary way by other workers. Instead of using *kT/h* as a prefactor, as does Hille, the prefactor is often chosen arbitrarily to fit the data.

The sad reality is that most many molecular biologists believe it is acceptable to 'scale' a theory, without realizing the absurdity of this view. What is special about multiplication by a constant? Why not allow arbitrary addition, or exponentiation, or use of some other function?

Obviously, traditional barrier models can fit experimental data taken in one solution if the prefactor is chosen arbitrarily.

If one wishes simply to fit equations to data, scaling or almost any other mathematical manipulation is fine, as long as it fits the data and provides a unique result. But uniqueness is a real issue when arbitrary prefactors are used. Choosing different conditions to determine the prefactor would produce different estimates of barrier height and thus different physical conclusions. What is surprising is that barrier models rarely can fit the current measured over a wide range of potentials and concentrations *even if used with an arbitrary prefactor* (that is held constant over the range of potentials and concentrations).

***Physical theories cannot be fiddled***. This misapplication of barrier models is a symptom of a general problem. Barrier models are physical theories, with parameters and functions that are supposed to mean something. They are not supposed to represent the arbitrary fits of functions, nor are they used that way. Barrier models are widely used precisely for the purpose of linking physical properties of molecules (e.g., their structure) with experimental measurements of current. They therefore cannot be arbitrarily manipulated or fiddled.

Unless we are vitalists, physical theories must be used as given us by physical scientists, who have gone to no small effort to derive and test them. We cannot take physical theories and multiply them by constants (or add constants or change them in any arbitrary way). We can of course behave as physical scientists and make up new theories or approximations, appropriate for our systems, but then they must be derived, simulated, and tested with the discipline of physical science, in papers refereed and published in the journals of those sciences.

Barrier models of channels and biochemical kinetics have certainly not been tested in this way, by derivation, simulation, or independent experimental check. Indeed, one characteristic of the literature on 'Eyring' models of enzymes, and barrier models of channels, is the nearly complete lack of references to the physical literature, certainly of the last 30 years, despite the enormous amount of work in this field, well over 700 papers (Fleming & Hänggi, 1993; Hänggi et al., 1990). What has happened is simply a bad turn in the history of science, caused (in my personal view) by the enormous financial and professional pressures for productivity that have led to an oversight, an ignoring of the relevant physical literature by molecular biologists.

It seems clear that traditional barrier models of permeation must be replaced. How to do that is of course another question altogether.

***Towards the future.*** The obvious candidate to replace verbal models is molecular dynamics, the direct computation of the motion of the atoms of channels.

The difficulties with molecular dynamics have been discussed before by me (Eisenberg, 1996a; Eisenberg, 1996b) following many others (Allen & Tildesley, 1987; Frenkel & Smit, 1996; Gaspard, 1998; Ott, 1997; Ott, Sauer & Yorke, 1994; Rapoport, 1997), but they too need reiteration because the evident visual appeal of molecular cinema in atomic detail tends to overwhelm one's critical faculties; certainly mine.

The fundamental difficulties of molecular dynamics are:

1) Present calculations are restricted to equilibrium. Thus, current cannot be predicted: attempts to predict a current that has already been assumed to be zero are not self-consistent, not unique, and frankly don't make sense. Much work is going on to remove these restrictions, in systems without electric charge (Heffelfinger & Swol, 1994; MacElroy, 1994), and no doubt that work will eventually succeed, but as of now, no one has published simulations of the dynamics of systems involving charged particles away from equilibrium.

2) Present calculations of the molecular dynamics of proteins rarely include ions in the surrounding solution. Since the properties of both proteins and channels are known experimentally to depend on the presence, concentration, and type of ion in the bath, simulations that do not contain ions there pose certain difficulties. *Proteins need ions, and so simulations of proteins need them, too*, particularly the simulations of protein folding and drug binding that are performed so often because of their evident importance.

   Simulations have not included ions because the systems simulated have been too small to define a definite concentration (with reasonable fluctuations) and because no one has known how to calculate the electric field when concentrations of ions are present. I believe systems must be large enough to define a concentration; and ions must be treated realistically enough to reproduce the relevant experimental properties of bulk solutions (i.e., the activity and conductivity actually measured in those solutions). Otherwise the simulations cannot hope to deal with a real biological system embedded in such solutions. Real biological systems are known experimentally to depend sensitively on the properties of the solution and so the solution must be included realistically in simulations of biological systems.

3) Simulations must extend long enough in time to calculate phenomena of biological interest. If the phenomena take seconds, it seems likely that the simulation must extend to seconds. If the simulation does not extend this long, the simulation must be extended artificially, either by argument or theory, and then has lost most of the advantages claimed for molecular dynamics. If the phenomena is found to occur more quickly in simulations than in life, the simulation is giving results different from the experiment, and it is unlikely to be useful.

4) Simulations must correctly sample the system being modeled. Since only a tiny subset of possible trajectories are computed, one must be sure that this subset represents the trajectories that are biologically and experimentally relevant. One must be sure the trajectories do not fall into one isolated domain, near one local minimum, which happens not to produce the biological behavior of interest.

The importance of this problem must be emphasized. The equations of molecular dynamics exhibit all the symptoms of chaotic mechanical systems. It is easy to verify that after a few picoseconds trajectories diverge exponentially and are exponentially sensitive to the choice of initial conditions. It is not true that the average properties of trajectories of chaotic systems reproduce the thermodynamic properties observed from such systems, because it is common to find that trajectories computed from chaotic systems are trapped in particular unrepresentative regions of phase space. Trajectories computed in such systems rarely sample the same space that real trajectories sample. Thus, the simulations of molecular dynamics are likely to miss many of the domains of biological interest.

Difficulties of this sort are, of course, not unique to biological systems and they are the main reasons that theories of lower resolution than molecular dynamics are so widely used in the physical sciences. Indeed, the great tradition of physics is to construct the theory of minimal complexity that accounts for the detail of experimental results, using as much atomic resolution as necessary, but not more. The work of John Bardeen (see April, 1992, issue of *Physics Today*, viz, Vol. 45(4), 1-136) illustrates this approach, and it is the approach adopted in *PNP*.

The problem is, of course, how to construct such a lower resolution theory and how to use it. One such theory, the (nonlinear) Poisson-Boltzmann (*PBn*) theory of proteins has had much success (Davis & McCammon, 1990; Forsten et al., 1994; Honig & Nichols, 1995) and has recently been extended to channels (Weetman, Goldman & Gray, 1997). The difficulty here is that channels do little at equilibrium, and thus calculations confined to equilibrium can not show what channels do. These ideas have been said before in an abstract way (e.g., Eisenberg, 1998b; Eisenberg, 1996b) but it seems that an argument by example is needed as well.

Of course, close enough to equilibrium, a conductance can be determined (in terms of the structure of the channel and physical parameters) from a quasi-equilibrium theory. If the reversal potential of the linear *I-V* characteristic can also be determined by the theory (in terms of the structure of the channel and physical parameters), then the quasi-equilibrium description is complete and fully satisfactory, for our purposes. As long as the conductance and reversal potentials are enough to describe the channel (over a range of concentrations and potentials); the near equilibrium is useful. Most open channel *I-V* curves are not that linear, however, and expressions for the reversal potential, and its variation with concentrations are not easily derived from quasi-equilibrium theories. In my opinion, near equilibrium descriptions are rarely useful, and never (to the best of my knowledge) over a reasonable range of experimental conditions, including asymmetrical solutions, with unequal concentrations of permeating ions on the sides of the channel.

***Why equilibrium calculations cannot predict or approximate flux*.** Calculations at equilibrium using (for example) *PBn* predict a potential profile through the channel and the accompanying profile of concentration (i.e., the probability of location of ions). These calculations have to be done under conditions of equilibrium, i.e., with bath concentrations and *trans*membrane potentials that produce zero flux of each ionic species, because *PBn* assumes equilibrium. If nonequilibrium conditions are substituted into the *PBn* equations, e.g., unequal concentrations with zero *trans*membrane potential, the equations cannot be solved, because flux must occur and be described by a nonzero number, but the theory assumes flux is zero. Indeed, no variable describing flux appears in the equations. If a computer program implementing *PBn* appears to give a result when run under nonequilibrium conditions, it must be incorrectly programmed, or it must not have converged.

The physical reason for these difficulties is that the potential and concentration profiles within the channel change when bath concentrations and/or *trans*membrane potentials are moved from their equilibrium values and produce current flow. The profiles have to be different, of course; otherwise, why would the current flow?

Specifically, imagine a perfectly selective $Na^+$ channel with 100 mM NaCl on one (left or in)side and 10 mM NaCl on the other (right or out)side. When the electrical potential is the Nernst potential, here around –60 mV, there will be no current flow, and *PBn* can be used to compute the potential profile $\varphi(x)$.

However, if the concentration of NaCl on the left side is changed to any other value, say for example 10 mM NaCl, and the electrical potential is not changed, i.e., it remains at –60 mV, the potential profile $\varphi(x)$ clearly must change (because the average contents of the channel must change, i.e., shielding changes, and this must change the potential profile $\varphi(x)$). *PBn* cannot calculate this new potential profile, because *PBn* assumes equilibrium, i.e., no flux of any species. Indeed, it does not contain a variable for flux. Thus, it must predict zero flux even when the boundary conditions guarantee that flux must flow.

This is the essential point and this is what we mean when we say that *PBn* or other equilibrium theories or simulations cannot be used to predict *I-V* curves.

The question then arises whether an equilibrium calculation might approximate the current that flows in nonequilibrium situations. It is easy to see that this cannot be so in the great majority of cases, although in special cases it might be possible (Dieckmann et al., 1999). Consider what would happen if in the previous example, the concentration of ions on both sides of the channel are raised but the channel is kept at equilibrium. For example, imagine the $Na^+$ channel with 200 mM NaCl on one (left or in)side and 20 mM NaCl on the other (right or out)side, still with a *trans*membrane potential of –60 mV. It is obvious that the concentration of ions at the ends of the channel will be quite different from that present when the channel is surrounded by 100 mM NaCl and 10 mM NaCl (at the same *trans*membrane potential). It is obvious that if a nonequilibrium situation were used with say 200 mM NaCl on the left or inside, then the concentration of $Na^+$ inside the channel on that side would be more or less what it is in the equilibrium case with 200 mM NaCl on the same side. So we can use the effect of concentration on equilibrium properties to (crudely) estimate its effect on nonequilibrium properties. In this way, it is clear that changing the concentration can have a large effect on the concentration at the end of channels and thus on their properties. Part of the reason is because of the Ohm's law effect (i.e., the current flow is accompanied by a separation of charge and thus a change in potential); but part is also simply because the concentrations of ions in nonequilibrium situations that produce flux are different from the concentrations present in equilibrium situations that do not produce flux.

Another way to see this is to consider two cases of the same channel (i.e., perfectly selective) with equal concentrations of salt on both sides, but different electrical potentials.

For example, compare 100 mM NaCl || 100 mM NaCl and 0 mV membrane potential and 100 mM NaCl || 100 mM NaCl and 100 mV membrane potential

*PBn* can predict the potential profile $\varphi(x)$ in the first place. Clearly *PBn* cannot predict the potential profile $\varphi(x)$) in the second place. Can *PBn* approximate the effect of the potential change? The size of the effect can be estimated by simply looking at the change in membrane potential. The membrane potential in the channel near the bath will change more or less as much as the bath potential changes. Thus, one would expect even a 10 mV change in *trans*membrane potential to have a large nonlinear effect on the potential profile $\varphi(x)$, because 10 mV is a substantial fraction of $kT/\boldsymbol{e}$ which is some 25 mV at room temperature. And calculations with *PNP* indeed show such a substantial effect. Another way to estimate the size of the effect is to use Ohm's law, and determine how much change in potential accompanies the currents that are measured experimentally. Again the potential changes within the channel are nearly always a substantial fraction of $kT/\boldsymbol{e}$ and so have substantial nonlinear effects.

The crucial point is that the potential profile $\varphi(x)$ is *not* just a function of the channel at hand (i.e., its structure and fixed charge, etc) but also a function of the average concentration of ions in the baths, in the channel's pore, and of the *trans*membrane potential.

How could it not be? If the potential profile were not a significant function of bath concentration and *trans*membrane potential, the free energy for moving an ion through a channel would be independent of the average concentration of ions in the baths, in the channel's pore, and of the transmembrane potential.

These variables—average concentration of ions in the baths, in the channel's pore, and the *trans*membrane potential—are substantially different in equilibrium and nonequilibrium situations. Thus, equilibrium calculations do not approximate the nonequilibrium situation in which channels function. Or to put it baldly, Ohm's law and Fick's law (or their equivalent) are needed to describe open ionic channels and those laws do not appear in, nor can they be derived from equilibrium calculations.

***Appropriate Models Now and in the Future***. The natural question then arises, what nonequilibrium models should be used to describe ion permeation? What can be used, given that direct simulation by molecular dynamics seems impractical?

One possibility is the *PNP* theory presented here, but that theory has its limitations. As presented, *PNP* represents the one dimensional average of a full three dimensional theory. The equations of one dimensional *PNP* were not just written down, but rather were derived by a professional mathematician (Barcilon, 1992) in three distinct ways, two independent perturbation methods and one matched asymptotic expansion. All three methods were carefully checked in the refereeing process and all give the same result. The one dimensional equations can also be derived by direct spatial averaging (Chen et al., 1992). Thus, the one dimensional equations of *PNP* have been more strictly derived than most models of chemical kinetics (*loc. cit.*) in which one dimensional reaction paths are more or less written down (without derivation, and certainly without estimation of error terms) as approximations to behavior in a high dimensional phase space, (Chandrasekhar, 1943).

When the structure of a channel protein is not known, the one dimensional theory seems the appropriate model, at this level of resolution, and it has done reasonably well, so far (Chen et al., 1998a; Chen et al., 1999; Chen et al., 1997b; Chen et al., 1995; Nonner et al., 1998; Nonner & Eisenberg, 1998; Tang et al., 1997). But when the three dimensional structure is known, clearly one should use it, and that requires a three dimensional theory, even if the one dimensional theory is its well defined spatial average.

Fortunately, three dimensional versions of *PNP* are becoming available. As I write, two groups are computing them, using independent but related numerical methods: Kurnikova *et al.* 1999, has completed a lattice calculation of gramicidin and shown qualitative agreement with the measured properties of gramicidin. Hollerbach *et al.*, 1999, accurately predict the *I-V* relations of gramicidin directly from the structure, using an independently determined estimate of the diffusion coefficient of $Na^+$ in the channel. It is clear even from this early work that the three dimensional calculations are feasible and that they give results similar to the one dimensional average. But differences will no doubt emerge as the calculations are pursued, compared, and checked in a range of conditions and channels.

The *PNP* model suffers from at least three difficulties, even in three dimensions (Eisenberg, 1998a; Eisenberg, 1998b; Eisenberg, 1996b; Horn, 1998): it lacks chemistry and single filing, it lacks spatial resolution, and it does not deal with protein conformation changes thought to underlie gating.

Specific chemical interactions clearly occur in binding sites of proteins and it never occurred to me, or anyone else I know, that similar effects would be absent in channels: enzymes and channels are both proteins created by the same evolutionary process and subject to the same laws. It is a tautology (but also an oxymoron) to describe channels as enzymes (Eisenberg, 1990).

Classical models of channels are based on the idea that specific binding, essentially analogous to that found in enzymes, is the direct determinant of permeation: "more bound is more permeant." Specific binding of this type is not naturally described by an electrostatic mean field theory although (in the absence of covalent bond changes) the underlying forces are clearly electrostatic and can be described by Coulomb's law used in atomic detail (Feynman, 1939; Mehra, 1994, p. 71-79).

When binding is described in the simplest possible way (Chen, 1997), and combined with *PNP*, we (Nonner et al., 1998; Nonner & Eisenberg, 1998) were amazed to find complex behavior that cannot be at all described as 'more bound, more permeant'. (I hasten to add that in all these calculations mobility and diffusivity are kept constant (Nonner et al., 1998; Nonner & Eisenberg, 1998), have found that the bound ions produce such an enormous local potential that no ions of the same sign can move. Ions of the opposite sign (the unbound ions!) determine the reversal potential. Of course, this was a particular calculation and not a general analysis. Nonetheless, it is clear that the combination of binding and electrostatics will give results very different from those previously assumed.

***Predicting function from structure***. This same approach, *PNP* plus binding, can be used to predict the properties of channels in general. In particular, it can be used to predict the properties of the McK channel from *Streptomyces lividans* whose structure has recently been reported (Doyle et al., 1998): see structure 1BL8 of the Protein Data Bank at Brookhaven National Laboratory, Upton NY 11973-5000 (web site *http://pdb.pdb.bnl.gov*). Catacuzzeno, Nonner, Blum, and I (Catacuzzeno et al., 1999b) are building a *PNP* model from this structure, and it is already apparent that a wide range of the properties of $K^+$ channels are easily and naturally predicted in this way. For example, a one dimensional representation of the charge distribution gives a surprisingly good prediction of the *I-V* relations of $K^+$ channels, including the *AMFE*, if it is used with binding sites described by the *MSA* and Nonner's mean field flow model of single filing: the non-independent flux ratio arises naturally as do the quite complex and highly voltage dependent *I-V* relations, found in single channels in mixed divalent/monovalent solutions.

***Structural basis of selectivity, gating and modulation***. The structure of the McK channel is striking because it contains three elements, which seem likely to produce three of the more complex permeation properties of channels, namely, selectivity, gating and modulation.

The narrow pore on the extracellular side of the protein seems ideally suited to provide selectivity between ions, and our preliminary analysis (Catacuzzeno et al., 1999b) suggests that the *MSA* can account for the selectivity observed in other $K^+$ channels (see p. 1301 of Nonner & Eisenberg, 1998). Although the parameters of the *MSA* appropriate for the channel environment are not known directly, and must be adjusted to fit selectivity data, it seems clear that a treatment based on an *MSA* with fitted parameters is preferable to the alternative of biologists trying to create their own theory of selectivity independent of the work of physical chemists in bulk solution and has already been reasonably successful, as previously described (Catacuzzeno et al., 1999b).

The narrow pore of the McK channel empties into a roughly spherical central cavity which then joins another pore, on the cytoplasmic side of the channel. This in-pore, as I like to call it, is formed by ***nonpolar*** amino acids. The nonpolar lining of the in-pore was not expected: most workers have thought all pores would be lined with polar hydrophilic amino acids.

***Non-polar pores as modulation sites***. I suggested (at the Liblice Statistical Mechanics Conference, August, 1998: Nonner & Eisenberg, 1999) that ***nonpolar pores are likely to be the main sites of channel modulation.*** A nonpolar pore is a structure that seems designed to allow modulation of open channel current by nearby charges.

Electric charges near a nonpolar pore produce large changes in the potential profile inside a nonpolar pore. The nonpolar lining has low fixed charge and a low dielectric constant and thus provides little dielectric shielding and little permanent charge to swamp the effect of charged structures outside the pore. Anything that changes the charge distribution outside the nonpolar pore changes the potential profile inside it, thereby changing current. For example, binding of charged or polar molecules to nearby proteins would modulate current flow this way.

A polar lined pore is totally different: the fixed charge is large—as is the dielectric constant—and so charges outside the polar pore are both shielded and swamped, thus having little effect on the potential or permeation in the pore lumen. In contrast to nonpolar pores, the potential profile in polar pores is quite independent of nearby charges.

Interestingly, nonpolar pores are less likely to be well described by mean field theories. The fixed charge which helps the mean field dominate in polar pores (Blum, 1994; Bratko et al., 1991; Henderson et al., 1979) is hardly present. For those reasons, single file phenomena (etc.) are more likely to be important in the in-pore than in the selectivity filter of the McK channel, in my opinion.

***α-helices as gating particles.*** We have speculated (Nonner & Eisenberg, 1999) that the $\alpha$ helices of the McK channel might form the structural basis of the gating particles proposed for sometime (Hodgkin & Huxley, 1952) even though voltage dependent gating does not occur in this particular $K^+$ channel. These $\alpha$ helices seem ideally placed to be push rods, that move slightly in response to the electric potential difference between the two ends of the channel (i.e., the *trans*membrane potential) while being reasonably independent of the local electrical potential inside the cavity or pore itself. Thus, these $\alpha$ helices seem to have the properties long expected of the 'delayed rectifier' (Cole, 1947; Hodgkin et al., 1949; Hodgkin & Huxley, 1952), the voltage dependent system that opens and controls many $K^+$ channels. We imagine that this rectification is suppressed in the McK channel phenotype because of special properties of this particular channel that will not be found in classical voltage gated $K^+$ channels.

***Simulations with electrostatics and atomic resolution***. So far, the only theory able to fit a wide range of *I-V* curves is one dimensional *PNP*, a mean field theory without atomic resolutions (Note that the three dimensional version of *PNP* (Hollerbach et al., 1999; Kurnikova et al., 1999) has atomic resolution in space, but not time, and so does not deal correctly with single filing). It also remains a mean field theory that does not describe ions as spheres, even though it is solved in three dimensions. Everyone would prefer a theory of permeation with atomic resolution and single filing. The attraction, even seduction of atomic structures is felt by me, just as much as everyone else. As discussed previously, direct simulations of motion are not possible because of inherent limitations in present day methods of molecular dynamics. But perhaps one could simulate with lower time resolution, preserving atomic spatial resolution, while computing the electric field from the charges present

One way to do this is to represent atomic motion the way Einstein and Smoluchowski did, as Brownian motion, using Langevin equations. I shall not cite the extensive literature of this field, but just point out that apparently no one in the chemical literature has done simulations of Langevin motion in which the charged atoms create their own electric field (Coffey et al., 1996). That is to say, all the Langevin simulations of ionic solutions that I know about; e.g., (Canales & Sese, 1998) calculate the motion of atoms in a predefined profile of potential, and do not calculate the profile from the charges of the system. Self-consistent Langevin calculations have been done in the semiconductor literature (Arokianathan, Asenov & Davies, 1996), and have been shown to give useful and reliable results (Arokianathan, Asenov & Davies, 1998a; Arokianathan, Asenov & Davies, 1998b). It seems to me that such calculations are clearly needed to understand the atomic basis of permeation. A number of groups are working on this problem and results seem possible.

These self-consistent Langevin calculations promise to deal with the greatest surprise of *PNP*, the lack of clear sign of single file phenomena. Measurements of unidirectional flux through $K^+$ channels clearly show behavior different from that of the *PNP* theory. Chen & Eisenberg, 1993, discuss this issue at length, provide an introductory definition and analysis of unidirectional fluxes (see their Appendix) and provide extensive literature references. Measurements of unidirectional flux are made in ensembles of channels on time scales some $10^{11} - 10^{18}\times$ slower than the atomic collisions that produce single filing, i.e., in 10-100 seconds, compared to $10^{-16}$ –$10^{-12}$ seconds, and so allow plenty of time for complex unexpectedly correlated three dimensional trajectories, in which tracer ions might interchange positions and fluxes might behave in unexpected ways. Nonetheless, I certainly agree with the common wisdom that the ratios of fluxes observed are *prima facie* evidence for single filing. Nonner has in fact created a mean field flow model of single filing so it can be included self-consistently in *PNP2*.

Finally, it may be possible to do self-consistent molecular dynamics incorporating the electric field directly, using the methods of computational electronics (e.g., DAMOCLES, ; Hess, 1991; Hess et al., 1991; Kersch & Morokoff, 1995; Lundstrom, 1992; Reggiani, 1985; Venturi et al., 1989). While semiconductors are certainly not ionic solutions, or ionic channels, their holes and electrons are quasi-particles that move according to laws similar to those governing electrons (Assad & Lundstrom, 1998), on similar time scales, posing (if anything) more complex computational challenges (because the quasi-particles have finite lifetime, experience much more complex friction, and follow ballistic trajectories whose duration must be computed literally on the fly). The physics of ions in water is very different from the physics of quasi-particles in semiconductors, but the mathematical descriptions are quite similar, because the mathematics is an expression of conservation laws, more than anything else. Thus, the computational procedures of semiconductor physics should certainly be useful tools for studying ions and channels.

Simulations of the motion of holes and electrons are in many ways more advanced than those of ions; e.g., simulations of holes and electrons are always done away from equilibrium, in the presence of substantial fluxes, and they always include macroscopic electric fields resulting from bias potentials more than analogous to the *trans*membrane ' potentials of channels. (Otherwise, the simulations could not be used to design real transistors, which require bias potentials to function usefully.) Physicists familiar with these methods might find them revealing if applied to systems of ions and channels (Eisenberg, 1998a).

***Gating and Conformation Change***. *PNP* is a theory of the stationary properties of open channels, and as such is not concerned with gating or conformation change. Nonetheless, gating and conformation change are important determinants of channel function and it is natural to wonder how they can be treated in a self-consistent theory. The criticisms of barrier models of permeation do not directly apply to barrier models of gating, of course. There, it is clear that high barriers exist, because many, if not most gating processes follow exponential time courses at a given *trans*membrane potential (Magleby & Pallotta, 1983a; Magleby & Pallotta, 1983b). Nonetheless, one must wonder what prefactor is actually used in theories of activation that apply at a range membrane potentials (Schoppa & Sigworth, 1998; Zagotta, Hoshi & Aldrich, 1994). One must wonder whether the distressingly large number of states in those models reflects the complexities of the gating process or the inadequacy of the models and basis functions (exponentials) being used to describe it.

Recently, Sigg, Hong, and Bezanilla (1999) have described gating current as the result of the electrodiffusion of a gating particle over an assumed potential landscape, much as we once treated electrodiffusion of permeating ions moving over a potential landscape in the channel's pore (Barcilon et al., 1993; Cooper et al., 1988a; Cooper et al., 1988b; Eisenberg et al., 1995). If Sigg *et al.*, computed their potential profile from an assumed distribution of fixed charge, the motion of their gating particle would be described self-consistently, as we try to describe the motion of permeating ions in *PNP*. Of course, until a self-consistent theory of gating current is actually constructed, it cannot be clear that such a theory would work. Conceivably, it could fail to fit data reasonably well described already by Sigg *et al.*, .

The theories of gating just described are rather abstract, because the mechanism(s) of gating are not known; indeed, the structures involved are not known. One should point out, however, that there are some clues to the physical basis of the gating transitions that produce rectangular single channel currents. (Other forms of gating are likely to come from different structures and have a different physical basis, e.g. some surely arise from conformational changes and steric effects.) Rectangular currents are known to arise when ions jump onto binding sites in insulating regions of field effect transistors (Kirton & Uren, 1989) and similar currents occur in 'Coulomb blockade' (Grabert & Devoret, 1992). If a tiny (0.1%) time independent conformation change is put into a time dependent version of the *PNP* equations, currents are computed that turn on and off as channel currents do (Gardner, Jerome & Eisenberg, 1998). It will be interesting to see whether any of these physical analogies form a useful model of the opening and closing of single channels.

## Acknowledgement

Duan Chen and I have been exploring open channels now for many years. It has been a joy to share the journey with him. Many gifted collaborators and friends have helped us find the way: Victor Barcilon, John Tang, Kim Cooper, Mark Ratner, Ron Elber, Zeev Schuss, Joe Jerome, Chi-Wing Shu, Steve Traynelis, Eli Barkai, Jurg Rosenbusch, Tilman Schirmer, Raimund Dutzler, Jim Lear, Le Xu, Ashutosh Tripathy, Gerhard Meissner, Carl Gardner, Wolfgang Nonner, Dirk Gillespie, Uwe Hollerbach, and Lesser Blum, have been particularly important contributors (listed more or less in chronological order). I am most grateful for their interest and contributions.

The work was made possible by the steadfast and generous support of the NSF and DARPA (grant N65236-98-1-5409).